\documentclass[useAMS,usenatbib]{mn2e}

\usepackage[dvips]{graphicx}
\usepackage[]{txfonts}
\def\simless{\mathbin{\lower 3pt\hbox
{$\rlap{\raise 5pt\hbox{$\char'074$}}\mathchar"7218$}}}   %< or of order
\def\simmore{\mathbin{\lower 3pt\hbox
{$\rlap{\raise 5pt\hbox{$\char'076$}}\mathchar"7218$}}}   %> or of order
\newcommand{\be}{\begin{equation}}
\newcommand{\ee}{\end{equation}}
\title[Plasmoids in blazars: fast flares on a slow envelope]{Reconnection-driven plasmoids in blazars: fast flares on a slow envelope}

\author[Dimitrios Giannios]
{Dimitrios Giannios$^{1,2}$\thanks{E-mail: dgiannio@purdue.edu
 (DG)}  \\
$^{1}$Department of Physics, Purdue University, 525 Northwestern
Avenue, West Lafayette, IN 47907, USA\\
$^{2}$Max Planck Institute for Astrophysics, Box 1317, D-85741 Garching,
  Germany\\}

\begin{document}
\date{Received / Accepted}
\pagerange{\pageref{firstpage}--\pageref{lastpage}} \pubyear{2009}

\maketitle

\label{firstpage}

\begin{abstract}
TeV flares of duration of $\sim$10 minutes have been observed in several blazars.
The fast flaring requires compact regions in the jet that boost 
their emission towards the observer at an extreme Doppler factor of 
$\delta_{\rm em}\simmore$50. 
For $\sim$100 GeV photons to avoid annihilation in the broad line region of PKS 1222+216, 
the flares must come from large (pc) scales challenging most models proposed to explain them.
Here I elaborate on the magnetic reconnection minijet model for the blazar flaring,
focusing on the inherently time-dependent aspects of the process of magnetic reconnection. 
I argue that, for the physical conditions prevailing in blazar jets, the 
reconnection layer fragments leading to the formation a large number of plasmoids. 
Occasionally a plasmoid grows to become a large, ``monster'' plasmoid. 
I show that radiation emitted from the reconnection event can account for 
the observed ``envelope'' of $\sim$day-long blazar activity while radiation from 
monster plasmoids can power the fastest TeV flares. 
The model is applied to several blazars with observed fast flaring. 
The inferred distance of the dissipation zone from the black hole
and the typical size of the reconnection regions are $R_{\rm diss}\sim
0.3-1$ pc and $l' \simless 10^{16}$ cm, respectively. The required 
magnetization of the jet at this distance is modest: $\sigma\sim$ a few.
Such distance $R_{\rm diss}$ and reconnection size $l'$ are expected if
the jet contains field structures with size of the order of the
black-hole horizon.

\end{abstract} 
  

\section{Introduction} 
\label{intro}

Blazars are extremely bright and fast varying extragalactic sources observed throughout the
electromagnetic spectrum. Depending on the presence or absence of observed optical emission
lines they are classified as Flat Spectrum Radio Quasars (FSRQs) or BL Lac objects, 
respectively (Urry \& Padovani 1995). In either case, it is believed that the blazar emission
is powered by relativistic jets which emerge from supermassive 
black holes and beam their emission at our line of sight. 

Blazar variability on timescales ranging from hours to decades has been commonly 
observed at various wavelengths of the electromagnetic spectrum (see, e.g., B{\"o}ttcher 2007).
The recent discovery of blazar flaring on $\sim 5-10$ minute timescales
came as great surprise. Such extreme flaring has now been observed from several
objects both BL-Lacs and FSRQs [Markarian 421 (Gaidos et al. 1996); 
PKS 2155--304  (Aharonian et al. 2007; hereafter referred to
as PKS 2155), Markarian 501 (Albert et al. 2007; Mrk 501), and the 
prototype object of the class BL Lac (Arlen et al. 2013); and the FSRQ
source PKS 1222+216 (Aleksi\'c et al. 2011; hereafter PKS 1222)]. 
Fast-evolving TeV flares are, therefore, a generic feature of blazar activity.  

Ultra-fast flares pose several challenges to theoretical models for the blazar emission.
The observed timescale of variability is too short to originate
directly from the central engine. Modulations of the properties of the
plasma in the vicinity of the black hole are limited by causality arguments to be longer
that the light crossing time of the horizon $t_{\rm v}\simmore R_{\rm Sch}/c\simeq 10^4M_9$ sec,
where $R_{\rm Sch}=2GM_{\rm BH}/c^2$ and $M_{\rm BH}=10^9M_9M_{\odot}$.
The observed $\sim 10$-min-long flares are far more likely to originate  
from compact emitting regions that somehow form in the jet
 (Begelman et al. 2008; Giannios et al. 2009; Narayan \& Piran 2012).
Furthermore, for the TeV photons to escape the source in PKS 2155 
and Mrk 501 the emitting blob must have a Doppler boosting 
of $\delta\simmore 50-100$ towards the observer (Begelman et al. 
2008; Finke et al. 2008; Mastichiadis \& Moraitis 2008). This 
is much larger than the Lorentz factor $\Gamma_j\sim 10-20$
typically inferred for blazar jets from superluminal motions
(see, e.g., Savolainen et al. 2010). Moreover, the $\gamma$-ray flaring from PKS 1222
directly constrains the location of the emitting region.
For $\simmore 100$ GeV photons to escape the {\it observed} broad line
region of the FSRQ,
the emitting region must be located at scales $\simmore 0.5$ pc
(Tavecchio et al. 2011; Nalewajko et al. 2012). This constraint is practically independent
of the assumed geometry of the broad-line region (Tavecchio \&
Ghisellini 2012).
Given the large dissipation distance and the
large inferred energy density at the source, the 
intense flaring from PKS 1222 implies an unrealistically large jet power 
unless the emitting material is, again, strongly boosting its emission 
with $\delta\simmore 50$ (Nalewajko et al. 2012)\footnote{For 
PKS 1222 a lower $\delta\sim 20$
is allowed {\it if} the distribution of energetic particles is extremely focused  
in the rest frame of the emitting material; see Nalewajko et al. (2012).}. 
One final clue for the origin of the fast flaring is that
it is observed on top of an envelope of longer $\sim$day-long flares. 
During the fast flaring the flux increases by a factor of $\sim$ 
a few with respect that of the envelope (Aharonian et al. 2007; Albert
et al. 2007).  

A large number of theoretical interpretations have been put forward
to explain the fast flares in {\it individual} sources. 
Fast beams of particles at the light cylinder (Ghisellini \& Tavecchio
2008) or the interaction of the jet with a red giant star (Barkov et al. 2012)
are some of them. Rarefaction waves in magnetized shells (Lyutikov \&
Lister 2010), relativistic turbulence in the jet (Narayan \& Piran 2012) or 
reconnection-driven minijets (Giannios et al. 2009) 
may also be responsible for the fast flares.
I focus here on the latter possibility.  
   
In the MHD jet paradigm (Blandford \& Payne 1982) 
the jet is expected to emerge from the central engine 
in the form of a Poynting-flux dominated flow. 
If the magnetic field configuration is appropriate
after the jet acceleration and collimation phases, 
magnetic reconnection can effectively dissipate 
magnetic energy in the flow. As pointed out in 
Giannios et al. (2009; 2010; Nalewajko et al. 2011) 
magnetic reconnection dissipates energy in compact regions 
characterized by fast motions within the jet, i.e., the radiating plasma
can move faster than the jet material on average. The extreme Doppler 
boosting of the emitting region and its small size can naturally
account for the fast-evolving flares observed in blazars. 

The reconnection minijet model is, however,  based on a {\it steady} reconnection picture.
Both observations and recent advances in reconnection theory reveal that reconnection
is a highly time-dependent and dynamical process. Time dependent aspects of reconnection
turn out to be critical in understanding the multiple observed timescales related to blazar flaring.
The goal of this work it twofold: (i) to relax the steady-state
assumptions of the reconnection model for blazar flaring 
and (ii) confront the model against all the available observational constraints.

In the Sec. 2 I summarize some of the recent observational and theoretical progress in understanding
time-dependent aspects of magnetic reconnection. In Sec. 3 this knowledge is 
applied to a blazar jets predicting the relevant timescales and energetics
of flaring. The model is applied to specific sources in Sec.~4. I conclude in Sec.~5.

\section{Magnetic reconnection: a dynamical process}

In this Section I summarize recent progress in magnetic reconnection
theory that is relevant to the blazar jet application presented here.
Reconnection is the process of fast release of magnetic energy during
a topological rearrangement of magnetic field lines. It takes place
when magnetic field lines of opposite polarity are coming together 
towards the reconnection plane ($x-y$ plane in fig.~1) 
and annihilate, liberating magnetic energy that
heats the plasma and accelerates particles.
The large scale $l$ of the reconnection region is determined
by the distance over which the magnetic field strength drops
by a factor of $\sim$2 (along the $y$ direction). 
The magnetic pressure gradient and also magnetic tension
along the $y$ direction result in the bulk acceleration of 
the reconnected material to the 
Alfv\'en speed $V_A$ of the upstream fluid. 
The fast outflow in the downstream allows for fresh magnetized fluid 
to enter the region and reconnect.

Observationally reconnection has been extensively studied
during solar flares and in Earth's magnetotail. Laboratory
experiments complement these studies in a controlled environment. 
A richness of processes that take place on very different 
timescales have been revealed by these works
(see, e.g., Aschwanden 2002). A characteristic long timescale of the process
is the global reconnection timescale  $t_{\rm rec} \sim l/\epsilon
V_A$ over which the magnetic energy stored in a region of typical 
scale\footnote{For simplicity and throughout this paper I will 
assume that the reconnection region
has the same characteristic scale $l$ in all directions.} $l$ 
is released. Here $\epsilon$
parametrizes the reconnection speed; with $\epsilon \sim 0.1$
been a typical observationally inferred value.
{\it Besides the global reconnection timescale,
much shorter timescale variability and eruptive events are evident
both observationally and experimentally 
highlighting the very dynamical nature of the process}
(see e.g., Lin et al. 2005; Park et al. 2006; Karlick\'y \& Kliem 2010).
For instance a solar flare of typical duration of $\sim$10 min can
show strong variability on $\sim$s timescale (see, e.g., Karlick\'y \& Kliem 2010).  

For some time the reconnection theory has been dominated by steady-state models
(Sweet 1958; Parker 1957, Petschek 1964). They provide intuition on how {\it average} 
properties such as the reconnection speed, the outflow speed and 
temperature of the reconnected fluid depend on parameters.
Steady state models assume a continuous inflow of plasma in the 
reconnection region and a smooth outflow. As such they
cannot account for the erratic behavior observed at the current sheet.

A distinctly different picture has emerged from recent theoretical studies of
magnetic reconnection. When the resistivity $\eta$ is sufficiently low, e.g., 
the corresponding Lundquist number $S=V_Al/\eta\gg S_c=10^4$ 
(as expected in most Astrophysical environments) the reconnection current
sheet is formally predicted by the Sweet-Parker theory to be extremely thin, with thickness $\delta/l=S^{-1/2}\ll 
S_c^{-1/2}\sim 0.01$. Very thin current sheets suffer from tearing instabilities
that lead to their fragmentation to a large number of plasmoids separated by smaller current sheets
(Loureiro et al. 2007; Lapenta 2008; Daughton et al. 2009; Loureiro et
al. 2009; Samtaney et al. 2009; Bhattachackarjee et al. 2009;
Loureiro et al. 2012b). As a
result the reconnection process is fast and resistivity-independent.
Plasmoids grow fast through mergers and 
leave the reconnection at a speed comparable to the Alfv\'en speed 
of the upstream plasma (Uzdensky et al. 2010; Loureiro et al. 2012a). The typical plasmoid
forms away from the reconnection center (the, so-called  x-point) at $y\sim l$ 
and grows up to a characteristic size $R_{\rm p}\ll l$.
Plasmoids that form fairly close to the reconnection center
have, however, more time available to merge and grow on their way out of the 
reconnection region. These
plasmoids  undergo significant (exponential) growth to reach
macroscopic scale. They are referred to as ``monster'' plasmoids (Uzdensky et al. 2010). 
Their size reaches a scale $R_p \simeq S_c^{-1/4}l\equiv f l\simeq 0.1
l$, i.e., approaching the global reconnection scale.  
The growth of the size
of a plasmoid is exponential $\propto$e$^{t/t_A}$, where $t_A=l/V_A$. 
The mass doubling of the plasmoid takes place on a timescale 
$\sim t_A$ with the plasmoid emerging from the reconnection region
at the Alfv\'en speed $V_A$. 
Resistive MHD simulations support this theoretical picture
(see, e.g., Bhattachackarjee et al. 2009; Loureiro et al. 2012a).

The monster plasmoids consist of energetic particles that have undergone acceleration
in the secondary current sheets. Additional acceleration of particles takes place
during the merging process of the plasmoids (Zenitani \& Hoshino 2001;
2005; Drake et al. 2006, Sironi \& Spitkovsky 2011;
although the energy spectrum of the particles depends on the details and
is still topic of investigation). As I demonstrate below, 
the macroscopic scale of the monster plasmoids, their short growth 
timescale, and fast motion and energetic particles that they contain
make them natural candidates for powering the ultra-fast blazar flares.

%-----------------------------------------------------------------  
\begin{figure}
\resizebox{\hsize}{!}{\includegraphics[]{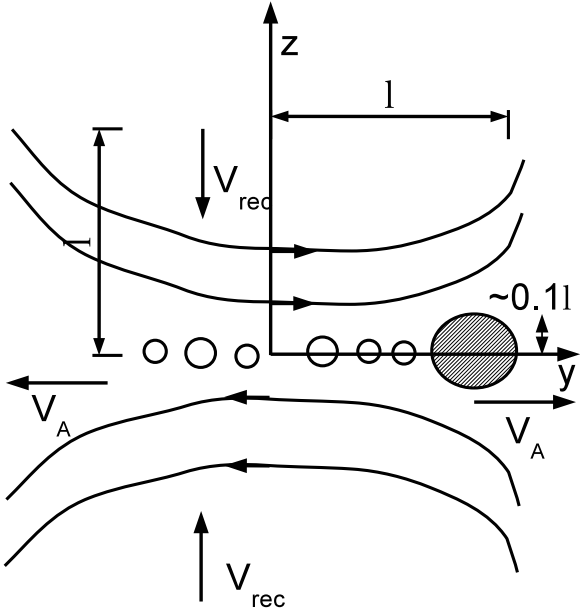}}
\caption[] {Sketch of the magnetic reconnection region with a large scale
$l$. Magnetic field lines of opposite polarity annihilate at the $x-y$ 
plane with speed $v_{\rm rec}$, i.e., the reconnection proceeds on 
a global timescale $t_{\rm rec}=l/\epsilon c$. The physical conditions
relevant to AGN jets, the current layer is expected to fragment 
to a large number of sub-layers separated by plasmoids (Loureiro et al. 2007). 
The  plasmoids leave the reconnection region with the Alfv\'en speed $V_A$
powering the ``envelope'' emission.
Occasionally, plasmoids grow to become ``monster'' plasmoids (shaded blob)
with scale $fl\sim 0.1 l$ giving rise to powerful, fast-evolving flares
of duration $t_{\rm flare}\ll t_{\rm rec}$.  
  
\label{fig1}}
\end{figure}
%----------------------------------------------------------------- 

\section{Application to blazar flaring}
\label{}

In the MHD-driving paradigm for jets (e.g., Blandford \& Payne 1982),
it is postulated that jets emerge from the central engine in the form
of Poynting-flux dominated flows. 
Further out the flow converts part of its magnetic energy into
kinetic. At a (both theoretically and observationally very uncertain) 
distance $R_{\rm diss}$, the blazar emission emerges. 
If the magnetic field configuration is appropriate, magnetic reconnection can
effectively dissipate magnetic energy in the flow and power the blazar
emission. I assume that substantial magnetic energy release takes place in reconnection
regions of characteristic scale $l'$.\footnote{Hereafter primed quantities are measured in 
the rest frame of the jet while double-primed quantities in the rest frame of
a plasmoid.}  The magnetization $\sigma\equiv B'^2/4\pi\rho c^2$ of
the jet at the dissipation region is assumed to be $\sigma\simmore 1$. As a
result, the Alfv\'en four velocity and, correspondingly, that of the reconnection outflows is expected to
be moderately relativistic $u_{\rm out}\simeq u_A=\sqrt{\sigma}$ (Lyubarsky 2005).

The location of $R_{\rm diss}$ and the scale $l'$ are highly model dependent.
The trigger of magnetic dissipation may be instabilities that develop in the jet. 
Even if the jet is launched by an axisymmetric magnetic field configuration, non-axisymmetric 
instabilities can introduce smaller scale field
structures. Current-Driven Instabilities (CDIs) are likely to be the most relevant in strongly magnetized jets
(Eichler 1993; Begelman 1998; Giannios \& Spruit 2006). The observational indications that
the jet opening angle is related to the jet Lorentz factor through the relation
$\theta_{j}\Gamma_j\sim 0.2$ (Pushkarev et al. 2009) implies 
that causal contact is established in the transverse direction of a
high-$\sigma$ jet. Under these conditions, CDIs can potentially grow as soon as the jet 
develops a dominant toroidal field component. CDIs are non-axisymmetric 
instabilities that reorganize the field configuration. 
In the non-linear stages of their development small
scale field reversals may be induced in the jet allowing for energy
extraction through reconnection (Moll 2009). The non-linear stages of CDIs are,
however, poorly understood making it hard to predict 
the distance at which they develop or the characteristic scales of the reconnection layers. 

An interesting alternative is that the magnetic field is not axisymmetric at the launching region.
The jet contains, instead, small-scale field structures imprinted from the
central engine (Romanova \& Lovelace 1997, Levinson \& van Putten 1997, 
Giannios 2010, McKinney \& Uzdensky 2012). 
Such configurations can introduce field reversals in the jet of the order
of the size of the black hole horizon $R_{\rm sch}= 3\times
10^{14}M_9$ cm (in the lab frame). 
The scale of the field reversal in the rest frame of the jet is $l'\simeq \Gamma_jR_{\rm Sch}
\simeq 6\times 10^{15}\Gamma_{\rm j,20}M_9$ cm. In such configuration,
substantial dissipation takes place when
the reconnection timescale $t_{\rm rec}=l'/\epsilon c=\Gamma_j R_{\rm Sch}/\epsilon c$ 
becomes comparable to the expansion timescale of the jet $t_{\rm exp}=R_{\rm diss}/\Gamma_jc$, 
i.e., at a distance 
\be 
R_{\rm diss}\simeq \Gamma_j^2R_{\rm Sch}/\epsilon= 1.2 \times 10^{18}M_9
\Gamma_{j,20}^2\epsilon_{-1}^{-1} \rm \quad cm. 
\ee
In the following, and for more concrete estimates, I adopt $R_{\rm diss}$ given by eq.~(1) and
$l'=\Gamma_{\rm j}R_{\rm Sch}$ as motivated by the proceeding discussion.
The model presented here can, however, be applied to any choice of the
parameters $R_{\rm diss}$ and $l'$.

%-----------------------------------------------------------------  
\begin{figure}
\resizebox{\hsize}{!}{\includegraphics[]{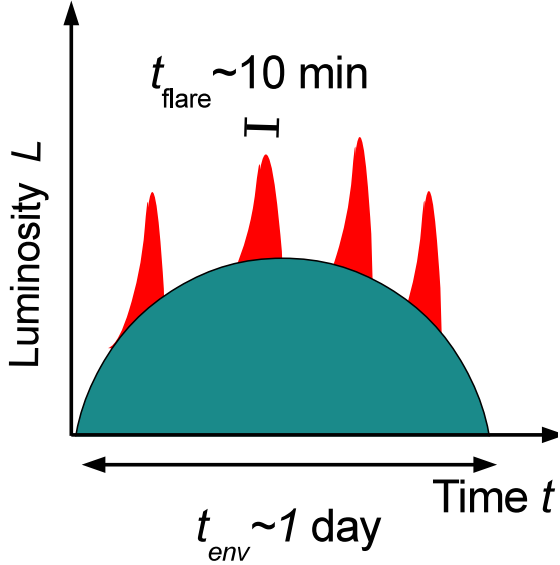}}
\caption[] {Sketch of the envelope-flare structure of the emission
from a reconnection layer. The envelope duration corresponds to that
of the reconnection event: $t_{\rm env}= l'/\Gamma_j\epsilon c$.
Monster plasmoids power fast flares which show exponential rise
and last for $t_{\rm flare}=0.1l'/\delta_{\rm p} c$.   
For an envelope of $\sim 1$day blazar flaring the model predicts that 
monster plasmoids result in $\sim 10$-minute flares. 
\label{fig1}}
\end{figure}
%----------------------------------------------------------------- 

\subsection{Fast flares from monster plasmoids}

For the physical conditions prevailing at the reconnection region relevant
to AGN jets, the current sheet is expected to fragment into a large number
of plasmoids while the reconnection process 
proceeds fast (see Appendix A for details).
Some plasmoids regularly grow into ``monster'' plasmoids, i.e., 
large magnetized blobs that contain energetic particles
freshly injected by the reconnection process (Uzdensky et al. 2010). The relativistic
motion of the plasmoids in the rest frame of the jet results in additional
beaming of their emission (i.e., beyond that induced by the jet motion).
When the layer's orientation is such that plasmoids beam their emission towards the
observer, powerful and fast evolving flares emerge.
{\it Here we focus on the characteristic observed timescales and luminosities
resulting from plasmoids that form in the reconnection region.} To this end
I assume that the dissipated energy is efficiently converted into radiation. 
In practice electrons are likely to be responsible for the emission 
(e.g. see Nalewajko et al. 2012)
so I, in effect, assume that a significant amount of the dissipated energy
is deposited to electrons which undergo fast radiative cooling. 
The latter assumption will be checked a posteriori. 
The former maybe justified by the efficient electron
acceleration by the electric field present at the  
current sheets but remains an assertion.  
This study can be trivially generalized by including the 
efficiency factor with which dissipated energy converts into radiation.

Consider a spherical blob (or plasmoid) emerging from the reconnection layer 
moving with the Alfv\'en speed of the reconnection upstream ($V_{\rm
  A}=\sqrt {\sigma/(1+\sigma)}c$), i.e, with a corresponding 
bulk Lorentz factor $\gamma_p=\sqrt{1+\sigma}\sim$ a few 
(measured in the jet rest frame) and of size $R_p''=fl'$. \footnote
{I treat the plasmoid as a sphere in its own rest frame (and not in
  the jet rest frame). It is unclear which approximation describes better
the reconnection plasmoids for relativistic reconnection. 
In the limit of modest relativistic motions of interest here
($\gamma_{\rm p} \sim$ a few), this distinction is not affecting the results presented here
by much.} 
The growth of the plasmoid in the reconnection layer is exponential
with time. The observed rise time
for the plasmoid emission is $t_{\rm rise,obs}\simeq R''_{\rm
  p}/\delta_p c$, where $\delta_{\rm p}$ is the Doppler boost of the plasmoid radiation towards the observer.
The plasmoid subsequently cools down and expands on a similar observed timescale $t_{\rm decline,obs}\sim
R''_{\rm p}/c\delta_{\rm p}$. We, therefore, define as the
variability timescale $t_v\equiv t_{\rm rise, obs}=fl'/c\delta_{\rm
  p}$. For field reversals imprinted from the central engine $l'\simeq
\Gamma_{\rm j} R_{\rm Sch}$ resulting in
\be
t_{\rm v}=\frac{f\Gamma_jR_{\rm Sch}}{\delta_{\rm p}c}=400
f_{-1}\Gamma_{\rm j, 20}M_9\delta_{p, 50}^{-1}\rm \quad s,
\ee
where $\delta_{\rm p}=50\delta_{\rm p,50}$, $f=0.1f_{-1}$, $\Gamma_{\rm
j}=20\Gamma_{\rm J,20}$.
Flaring on several minute timescale is therefore expected in this picture.

Consider a jet emerging from a supermassive black hole with
(isotropic equivalent) power $L_{\rm iso}$, opening angle $\theta_j$ and Lorentz factor 
$\Gamma_j$. We also assume that $\theta_j \Gamma_j=0.2$ as indicated by observations (Pushkarev et al. 2009). 
The typical bulk Lorentz factor of gamma-ray active blazars is $\Gamma_j\sim 10-20$ 
(Savolainen et al. 2010; Piner et al. 2012). The energy density at the
dissipation, or ``blazar'', zone is
\be
U'_{\rm j}=\frac{L_{\rm iso}}{4\pi (\theta_{\rm j} R_{\rm
    diss})^2\delta_{\rm j}^4c}=0.12\frac{L_{\rm
    iso,48}\epsilon_{-1}^{2}}{M_9^2\Gamma_{j,20}^2\delta_{\rm j,20}^4}
\quad \rm erg/cm^{3},
\ee
where the $L_{\rm iso}$ is normalized to $10^{48}$ erg/s and the
dissipation distance is given by eq (1). 

Pressure balance across 
the reconnection layer requires the energy density of the plasmoid to be similar to that
of the jet $U''_p\sim U'_j$.\footnote{The exact relation of $U''_p$ and $U'_j$ depends 
on the magnetic and heat content of the upstream and the plasmoid. 
For instance balancing the pressure of cold, strongly magnetized 
upstream $p_j=B'^2/8\pi=U'_j/2$ with that of an assumed 
relativistically hot plasmoid $p_p=U''_p/4$, we find $U''_{p}=2U'_j$.
The expression is slightly different if the magnetic field contributes
to the pressure of the plasmoid. The Assumption $U''_p\sim U'_j$ is expected
to hold within a factor of $\sim 2$ independently of these details.} 
Assuming efficient conversion of dissipated energy into radiation 
(assumption to be verified in  Sect. 4.3),
the rest-frame luminosity of the plasmoid is thus
$L''=U_{\rm p}''4\pi R_{\rm p}''^2c$. This luminosity can be converted to the
observed luminosity $L_{\rm p,obs}=\delta_p^4L''$. 
Because of the $R_{\rm p}''^2$ dependence of the luminosity it is clear that the largest 
``monster'' plasmoids (with $R_p''=fl'$, $f\simeq 0.1$) power the brightest flares.
Putting everything together, the observed luminosity of the plasmoid is
\be
L_{\rm p,obs}=10^{47}\frac{\epsilon_{-1}^2f_{-1}^2\delta_{p,50}^4L_{\rm iso,48}}{\delta_{j,20}^4}\quad \rm erg/s.
\label{lp}
\ee      

The Doppler factor of the plasmoid $\delta_{\rm p}$ depends on several parameters. It is 
related to $\Gamma_j$, $\gamma_p$, the angle of the plasmoid with
respect to the jet motion and the observer's angle of sight. For perfectly aligned
jet, plasmoid and observer $\delta_{\rm p} \simeq 4 \Gamma_{\rm
  j}\gamma_{\rm p}$. In, perhaps, more common situations where
the reconnection layer is at a large $\theta\sim \pi/2$ angle with respect to the jet propagation
(as seen in the jet rest) and fairly aligned with the observer (giving
powerful flares) $\delta_{\rm p} \sim 
\Gamma_{\rm j}\gamma_{\rm p}$. For the demonstrative arithmetic examples used here we adopt 
$\delta_{\rm p}=1.25 \Gamma_{\rm j}\gamma_{\rm p}=50\Gamma_{\rm j,20}\gamma_{\rm p,2}$.
One can see (see eq.~2) that powerful flares on a timescale of $\sim$10 min is possible even with
very modest relativistic motions within the jet\footnote {Narayan \& Piran (2012)
come to a different conclusion concerning the fastest flares expected by magnetic reconnection.
Their analysis assumes {\it steady} reconnection and is therefore applicable to the global reconnection
timescale (topic of next Section) but is not constraining the fast
flares powered by plasmoids.} $\gamma_{\rm p}\sim 2$.

\subsubsection{Ejection of multiple monster plasmoids}

During a reconnection event multiple monster plasmoids are expected to form. 
The seed of a monster plasmoid forms fairly close to the reconnection 
center region $y\ll l'$  and spends sufficient time in the reconnection
region to grow to a large size. 2D simulations (Loureiro et al. 2012a) 
indicate that monster plasmoids form every a few Alfv\'en times $t_A$
or at a rate of $\sim 0.3t_A^{-1}$. 
It appears likely that 2D simulations underestimate the rate of formation
of monster plasmoids. The actual rate may be higher when the 3D structure 
of the layer is considered. The x-point is in reality an elongated structure 
(along the x-axis of fig.~1) providing a larger physical region where
the seeds of monster plasmoids form. Monster 
plasmoids potentially form at a rate up to $l'/R'_{\rm p}\sim 10$ 
higher than that found in 2D studies. Clearly this question can only be 
answered high resolution, 3D restive MHD simulations. 
If monster plasmoids emerge at a rate $\sim (0.3-3)t_A^{-1}$,
some $(3-30)/\epsilon_{-1}$ plasmoids are expected 
from a single reconnection layer powering multiple flares. 
The observed properties of the monster plasmoids
are determined by the basic properties of the reconnection region that
generates them. To the extent that all monster plasmoids reach similar size of $\sim 0.1 l'$,
the model predicts a similar duration and brightness for this sequence
of fast flares. Smaller plasmoids $f<0.1$ can power even faster flares 
since $t_{\rm v}\propto f$ albeit of lower peak luminosity ($L_{\rm p}\propto f^2$).
A sketch of such pattern is shown in Fig. (2).

\subsection{The ``envelope emission'' from the reconnection region}

The bulk motion of a monster plasmoid is expected to be similar to the speed of other
structures (e.g. smaller plasmoids) leaving the reconnection region.
When the plasmoid emission is beamed towards the observer (powering a fast flare), the
overall emission from the current layer is also beamed by a similar factor. 
The emission from the layer forms a slower-evolving envelope.
In the following I calculate the timescale and luminosity of the
emission from the reconnection layer.

At the dissipation distance $R_{\rm diss}$, the reconnection proceeds
within the expansion time of the jet which is obseved to last for 
$t_{\rm exp,obs}\simeq R_{\rm diss}/\Gamma_j^2c$. Therefore, $t_{\rm exp,obs}$
corresponds to the observed duration of the envelope emission (using
also eq.~(1)): 
\be
t_{\rm env}=\frac{R_{\rm diss}}{\Gamma_j^2c}=
10^5\frac{M_9}{\epsilon_{-1}} \quad \rm s. 
\ee
The duration of the envelope emission is $\sim$days. 
Such timescale is characteristic of blazar flares (See next Section).

The (lab frame) energy available to power the envelope emission is $E_{\rm env}=U_{\rm j}2l'^3/\Gamma_{\rm j}$,
where $U_j=\Gamma_j^2U'_j$ is the energy density of the jet and $2l'^3/\Gamma_{\rm j}$
accounts for (lab frame) volume of the reconnection region that powers each minijet (see fig.~1).
The emitted luminosity of the reconnection region is $E_{\rm env}/t_{\rm env}$. It can be converted
into {\it observed} luminosity by accounting for beaming factor of the
emission $\sim \delta_p^2$:
\be
L_{\rm env,obs}\simeq 2\Gamma_{\rm j}^2\delta_{\rm p}^2l'^2U_{\rm
  j}'\epsilon c=3\times 10^{46}\frac{\Gamma_{\rm j,20}^2\delta_{\rm p,50}^2
\epsilon^3_{-1}L_{\rm iso, 48}}{\delta_{j,20}^4} \quad \rm erg/s. \label{lenv}
\ee

The envelope emission is quite bright.
Dividing eqs.~(\ref{lp}) and (\ref{lenv}), one arrives to a fairly simple expression
for the ratio of the plasmoid to envelope luminosities $L_p/L_{\rm
  env}\sim 3f_{-1}^2\delta_{\rm p,50}^2/\Gamma_{\rm j,20}^2\epsilon_{-1}$.
The luminosity contrast depends only on the Lorentz factor of the minijet in the rest frame of the
jet  $\gamma_{\rm p}\simeq \delta_{\rm p}/\Gamma_{\rm j}$, the size
of the plasmoid parametrized by $f$, 
and the reconnection rate $\epsilon$. As we discuss in the next
Sections, the luminosity ratio is observed to be of order unity constraining
$\delta_{\rm p,50}/\Gamma_{\rm j,20} \sim 1$ for $\epsilon\sim f\sim
0.1$. The ratio $\delta_{\rm p,50}/\Gamma_{\rm j,20}$ is determined by
the reconnection-induced bulk motions in the jet and points to
$\gamma_{\rm p}\sim 2$ or, equivalently, moderately magnetized jet
with $\sigma\sim $ a few.

So far I have considered the emission from a {\it single} reconnection
region which is beaming its emission towards the observer.  When the
reconnection scale is smaller than the cross section of the jet ($l'<R\theta_j$),
there may be as many as $\sim (R\theta_j/l')^3$ reconnection regions in the emitting zone. Even after corrected for
beaming of the emission from current sheets, up to $\sim (R\theta_j/l')^3/\gamma_{\rm p}^2$ 
reconnection regions may beam their emission towards the observer.
The overall amplitude of the envelope emission and the number of fast
flares are therefore enhanced by this factor\footnote 
{The opposite limit where $l'>R\theta_j$ is not physical since the reconnection region should fit
within the jet cross section: the condition $l'\simless R\theta_j$ must be satisfied.}. 
In this case, the contrast $L_{\rm p,obs}/L_{\rm env,obs}$ drops since more than one reconnection layers contribute
to the envelope emission.

\section{Application to observations}

In this section we examine how the wealth of information from fast TeV
flaring among blazars can be used to extract information
on the physical conditions of the emitting region and 
constrain the reconnection model. We first discuss observations
of several BL Lac objects and then FSRQ PKS 1222.

\subsection{Flaring BL Lacs}

Fast TeV flares have been observed in several BL Lac objects including
PKS 2155, Mrk 501, Mrk 421 and BL Lac itself. The variability timescale ranges from
a few to $\sim$10 minutes. Most detailed are the observations of
PKS 2155 which reveal an envelope emission of duration of hours that contains 
several $\sim5$-minute flares of comparable luminosity (Aharonian et
al. 2007). Mrk 501 also shows flaring with similar flare-to-envelope ratio. BL Lac, the prototype
source has also been observed to flare on 10 minutes suggesting that minute-flaring
is a generic property of blazars.

Here we apply the model to the, most constraining, PKS 2155 observations (see Aharonian et al. 2007).
The observed luminosity of the envelope and of the fast flares is  
$L_{\rm env}\sim L_{\rm ff}\sim 3 \times 10^{46}$ erg/s. The fast flares last for 
$t_v\sim 5$ min. When the observation started the envelope emission
was already on and lasted for $\sim$100 min. Narayan \& Piran (2012) 
use Bayesian analysis to estimate the mean duration of the envelope
emission of $t_{\rm env}\sim 2\times 10^4$ sec.
The isotropic equivalent jet power is uncertain but $L_{\rm iso}\sim
10^{48}$ erg/s appears a reasonable estimate given the {\it observed} luminosity 
of the source of up to $10^{47}$ erg/s and assuming an overall
radiative efficiency of $10\%$. 
A high Doppler factor  $\delta_{\rm p}\simmore 50$ of the emitting
material is required for the escape of the TeV radiation from
the soft radiation field of the jet without extensive pair creation (Begelman et al. 2008). 
Observations do not directly constrain where the emission takes place in this source.

The fact that $L_{\rm env}\sim L_{\rm ff}$ means that $\gamma_{\rm p}
\sim 2$. Setting $\delta_{\rm p}=50$ and $M=10^9M_{\odot}$ and the 
inferred $L_{\rm iso}$ in eqs. (2), (5), and (6), we derive
all fast flaring and envelope timescales and luminosities in good agreement
with the observed values. Moreover, PKS 2155 showed
(i) several fast flaring events of (ii) similar characteristic timescale and luminosity.
Multiple flares have a natural explanation within the reconnection model. 
They can be understood to come from different monster plasmoids that emerge 
from the same reconnection region. 

\subsection{TeV flares from PKS 1222}

The model can also be applied to the FSRQ PKS 1222. In this source 
the dissipation distance is robustly constrained to be $R_{\rm diss}\simmore 0.5$ pc
(Tavecchio et al. 2011; Nalewajko et al. 2012). 
If the emitting region is characterized by $\delta\sim \Gamma_j\sim 20$ the 
required luminosity of the jet is unrealistically high (Nalewajko et al. 2012).
A moderately higher $\delta\sim 50$ (in agreement to that inferred for PKS 2155) 
is, however, sufficient to relax the energetic requirements of the
jet and is adopted here. Around the epoch of the TeV flare there is an envelope of high
$\gamma$-ray activity. Fermi-LAT detected a flare of 
$L_{\rm env}\sim 10^{48}$ erg/s and duration of roughly $t_{\rm
  env}\sim 10^5$ sec (Tanaka et a. 2011; note, however, that {\it Fermi}
observations are not strictly simultaneous with the {\rm MAGIC} ones). 
The fast flares are observed with {\rm MAGIC} in the sub-TeV energy range:
$L_{\rm ff}\sim 10^{47}$ erg/s (with total luminosity possibly higher by 
$\sim$ a few to account for bolometric corrections in view of the steep
observed TeV spectrum). The flux evolved on a timescale of  
$t_v\sim 7$ min. The isotropic equivalent jet power is also 
uncertain but $L_{\rm iso}\sim 10^{49}$ erg/s appears a reasonable
estimate given that the observed disk emission is several $10^{46}$ erg/s 
and that the beamed observed radiative luminosity of the jet reaches $10^{48}$ erg/s. 

Setting $\delta_{\rm p}=50$, $L_{\rm iso}\sim 10^{49}$ erg/s, one can
verify that the observed duration of the envelope and of the fast flare are 
reproduced by the model. The same holds true for the flare luminosity.
The observed envelope emission is observed to be more luminous than the fast flare
by a factor of several (though weaker, the fast flare
is clearly observed with MAGIC because of its harder emission). 
Given the adopted parameters, eq.~(4) implies that a single reconnection
region has envelope luminosity $L_{\rm env}\sim L_{\rm ff}$ while 
the envelope was a factor of $\sim$several brighter in this source. 
Possibly several reconnection layers contribute to the envelope emission
simultaneously if $R\theta_j/l'\sim$ several, enhancing the ratio of the
luminosity of the envelope emission with respect to that of fast flares. 

Summarizing all blazar flares can be accounted for by little changes of the
physical parameters of the model. Typically I infer $\Gamma_{\rm
j}\sim 20$, $\gamma_{\rm p}\sim 2$ and the size of the reconnection 
region $l'\simless  10^{16}$ cm. The blazar zone is located at 
$R_{\rm diss}\sim (0.3-1)$ pc.

\subsection{Radiative mechanisms and particle cooling}

The energetic requirements for the fast flaring 
can become more stringent if the radiating particles
(assumed to be electrons) are not in the fast cooling regime.
Here we assume that the TeV emission is either result of synchrotron
self Compton (SSC) or external inverse Compton (EIC) and investigate
the electron energetics required to produce the observed $\sim 100$GeV- multi TeV emission  
and whether they are likely to radiate efficiently for the model parameters adopted
in the previous Section. In the end of the Section, I discuss the
expectation of X-ray flares as result of synchrotron radiation from
the same population of electrons. 

To assess whether the emitting particles ``cool fast'',
the expansion timescale of the plasmoid $t''_{\rm exp}=R''_{\rm p}/c=\delta_{\rm p} t_{\rm v}
=3\times 10^{4}\delta_{50}t_{\rm v,10}$ s
is to be compared with the radiative cooling timescale.
In the case of SSC emission from $e^-$ (or pairs)
with random Lorentz factor $\gamma_e$ in magnetic field $B''$, 
the characteristic energy is $e_{\rm SSC}\simeq 10^{-8}
\delta_{\rm p} \gamma_e^4B''$ eV. Depending on details of the
reconnection configuration (such as guide field strength), the
plasmoid can be magnetically or heat dominated\footnote{The fact that 
the jet is Poynting-flux dominated in this model does not necessarily mean that
the emitting region is also strongly magnetized. On the contrary, efficient 
reconnection may result in weakly-magnetized (heat-dominated) plasmoids.}. 
For simplicity, I parametrize the magnetic field strength in the plasmoid 
as $B''=(\epsilon_B4\pi U_p'')^{1/2}=0.7\epsilon_{\rm
  B,1/3}^{1/2}L_{\rm iso, 48}^{1/2}
\epsilon_{-1}M_9^{-1}\Gamma_{\rm j,20}^{-1}\delta_{\rm j,20}^{-2}$
Gauss. Setting $e_{\rm SSC}=100e_{\rm 100GeV}$ GeV, one finds for 
the required electron (random) Lorentz factor $\gamma_e=2\times 10^4e_{\rm 100GeV}^{1/4}
M_9^{1/4}\Gamma_{\rm j,20}^{1/4}\delta_{j,20}^{1/2}\epsilon_{-1}^{-1/4}\epsilon_{\rm
  B,1/3}^{-1/8}L^{-1/8}_{iso,48}\delta_{\rm p,50}^{-1/4}$. The SSC cooling
time scale is $t''_{\rm SSC}=5\times 10^{8}/(1+y)\gamma_eB''^2$ s or
\be
t''_{\rm SSC}\simeq 1.5\times
10^4\frac{3}{1+y}\frac{\delta_{\rm p,50}^{1/4}M_9^{7/4}\Gamma_{\rm
    j,20}^{7/4}\delta_{\rm j,20}^{7/2}}
{e_{\rm 100GeV}^{1/4}\epsilon_{-1}^{7/4}\epsilon_{\rm
    B,1/3}^{7/8}L_{\rm iso,48}^{7/8}}\quad \rm s,
\ee 
where $y\sim$ a few accounts for the ratio of the SSC to synchrotron power.

If a substantial external radiation field of energy density $U_{\rm rad}$
is present, it can contribute to the particle cooling through EIC. Assuming an isotropic
radiation field of characteristic energy $e_{\rm seed}$, the energy of the up-scattered photon 
is $e_{\rm EIC}\simeq \Gamma_{\rm p}\delta_{\rm p}\gamma_e^2e_{\rm seed}$ (for scattering in the Thomson limit).
Solving for the electron Lorentz factor: $\gamma_e\simeq 7\times 10^3(e_{\rm EIC, 100GeV}/
e_{\rm seed,1})^{1/2}\delta_{50}^{-1}$, where $e_{\rm seed}=1e_{\rm
  seed,1}$ eV and $\Gamma_{\rm p}\simeq \delta_{\rm p}$. 
The energy density of radiation in the rest frame of the blob is $U''_{\rm rad}\simeq 
\Gamma_{\rm p}^2U_{\rm rad}$. The EIC cooling timescale for such electron is
$t''_{\rm EIC}=2\times 10^7/\gamma_eU''_{\rm rad}$ s or
\be
t''_{\rm EIC}=\frac{1.4\times 10^4}{U_{\rm seed,-4}\delta_{\rm p,50}}\Big(\frac{e_{\rm seed,1}}{e_{\rm 100GeV}}\Big)^{1/2}\quad \rm s,
\ee  
for $U_{\rm seed}=10^{-4}U_{\rm seed,-4}$ erg cm$^{-3}$.

In the case of PKS 2155 no powerful ambient isotropic radiation field is evident. The plasmoid may
well propagate into a dense radiation field emerging from the large scale jet or
other reconnection regions. This depends however on uncertain details of the overall 
geometry (see Nalewajko et al. 2011 for various possible geometrical
configurations). On the other hand SSC emission 
has to be present. Setting the model parameters  to those relevant for
PKS 2155 (see previous Section), 
and $\epsilon_B=1/3$, $y=2$, I arrive at $t_{\rm SSC}\simless
10^4$ s (see eq.~5).
This timescale is, by a modest factor, shorter than the expansion time of the blob
$t''_{\rm exp}=1.5\times 10^4\delta_{50}t_{v,5}$ s indicating that efficient 
$\sim$TeV emission is plausible. The required random Lorentz factor of 
the emitting electrons is $\gamma_e\simmore 10^4$. 

One can derive a similar estimate for the SSC cooling timescale for the 
parameters relevant to PKS 1222 flaring, i.e. $t''_{\rm SSC}\simless t''_{\rm exp}\sim 2\times 10^4$s.
Another obvious possibility for emission from this source is EIC of
photons from the infrared torus.
With $U_{IR}\sim 10^{-4}$ erg cm$^{-3}$ and characteristic energy $e_{\rm IR}\sim$ 0.3 eV,
the EIC takes place in the Thomson regime (Nalewajko et al. 2012). 
The typical electron emitting $\sim 100$ GeV has 
$\gamma_e\sim 10^4$ with a cooling time scale of the particles of 
$t''_{EIC}\sim 8\times 10^3$s $\sim t''_{\rm SSC}\simless t''_{\rm exp}$.
  
In Summary the $\sim$TeV emitting electrons are characterized by a random $\gamma_e\sim 10^4$
and have a cooling timescale somewhat shorter than the expansion timescale of the blob
allowing for efficient TeV emission. The detailed spectrum necessarily depends on the
details of the particle distribution that are model
dependent. However, an equipartition argument (e.g. sharing of the
dissipated magnetic energy between electrons and protons) would 
give $\gamma_e \sim (m_p/2m_e)\sigma \simeq 3\times 10^3\sigma_3$ for the
electrons in the plasmoid, where $\sigma=3\sigma_3$ is the upstream
magnetization. Therefore, a modest particle acceleration 
above equipartition is sufficient to explain the observed emission. 

For the typical conditions inferred in the emitting region
($\gamma_e\simmore 10^4$, $B''\sim 1$ Gauss, $\delta_p\sim$ 50),
the synchrotron component naturally peaks in the soft X-ray band.
If SSC is the mechanism for the very high energy emission, the
synchrotron component may be quite powerful $L_{\rm syn}=L_{\gamma}/y$.
Fast X-ray flares, simultaneous to the very-high energy ones, 
are therefore quite likely in this model (see the Discussion section for
observational evidence for such flaring).

\section{Discussion/Conclusions}
\label{sec-conclusions}

The similarities in the variability patterns seen in 
several blazars (PKS 1221, PKS 2155, BL Lac,
Mrk 501) are striking: fast TeV flares on $\sim$minutes timescale
that appear on top of an 
envelope of enhanced gamma-ray activity that lasts
for hours or days. The similarities strongly indicate 
similar physical conditions at the emitting region:
large Doppler factor $\delta_{\rm p}\sim 50$ 
and a dissipation zone located at $\sim$pc distance from the black hole. 

\subsection{Models for fast blazar flaring}

Several suggestions have been put forward for the ultrafast blazar
flaring. Fast electron beams (with bulk $\gamma_e\sim 10^6$)
may develop along magnetic field lines close to the light cylinder
(i.e. within several $R_{\rm Sch}$, see, Ghisellini \& Tavecchio 2008).
The TeV flare, in this scenario, is result of the beam inverse Compton
scattering external radiation fields. This model fails to account 
for the large emission distance required by PKS 1222 (Aleksi\'c et
al. 2011). For a hadronic model applicable to the fast flares of
PKS 1222 see Dermer et al. (2012). In this model $\simmore 100$ TeV
neutrinos are predicted. Alternatively, a red giant may cross the jet (Barkov et al. 2012). 
The ram pressure of the jet strips the envelope of the star which 
consequently fragments. Emission from shocked stellar and/or jet plasma may
power blazar flares. While stellar-jet encounters are expected, 
any interaction region will necessarily move slower that the 
jet $\Gamma_{\rm int}\simless \Gamma_j$. The required Doppler
boost of the emitting region towards the observer $\delta\sim 50$ 
may therefore be hard to explain in this picture. Alternatively, a recollimation
shock on pc scales can help to focus the jet energy in a small region
inducing short timescale viability. However, non-axisymmetries
in the jet-external medium interaction will likely make the focusing
insufficient to explain the most extreme flares  
(Bromberg \& Levinson 2009). If the jet activity is sufficiently erratic,
the jet can be envisioned as magnetized shells separated by vacuum
(Lyutikov \& Lister 2010). Rarefaction waves of the front part of a shell
can reach a bulk Lorentz factor much higher than that of the jet on average. 
Fast flares may come from these rarefied regions.
Relativistic turbulence in the jet can also allow for emitters-blobs
moving with $\Gamma_b>>\Gamma_j$ to be responsible 
for intense and fast flares (Narayan \& Piran 2012).
For the turbulence not to be supersonic (or it would decay fast
by shocks) the jet must be Poynting flux dominated. The driver
and the region where the turbulence develops remain to be identified. 
Magnetic reconnection could drive the turbulent motions
(and turbulence can, in turn, enhance the reconnection rate;
Lazarian \& Vishniac 1999). In this case, however, it is quite likely
that the most powerful flares are directly related to the driver,
i.e., the reconnection process itself rather than the turbulent
motions.

\subsection{The reconnection model for fast flaring}

Here we have revisited the reconnection minijet model
for the fast flaring (Giannios et al. 2009; 2010). We focus on time-dependent aspects
that are naturally expected (and directly observed in laboratory experiments 
and solar system environment) in the reconnection process.
It is demonstrated that at least two timescales appear in the problem.
The longer one is associated with the time it takes for a magnetic
energy to be dissipated in the reconnection region and creates
an ``envelope'' of flaring activity that lasts for several hours to days.
Instabilities in the current sheet (e.g. secondary tearing
instability; Loureiro et al. 2007) result in erratic formation
of plasmoids that leave the reconnection region at relativistic
speed. The largest ones, ``monster'' plasmoids, can power the fast, 
$\simless 10$minute-long blazar flares. Several to tens of monster 
plasmoids can emerge from a single reconnection layer.  
The super-fast flaring may therefore not happen in isolation.
{\it A sequence of fast flares are expected to have similar timescale
set by the size of the reconnection layer as observed in PKS 2155.}
Verification of this trend of a sequence of flares in more sources 
and/or in other bands such  as X-rays (see below) 
would provide strong support for the model.

A virtue of the model is that it can be applied to all 
blazar sources with observed fast flaring for similar adopted parameters.
In this model, the dissipation of energy that powers the blazar
radiation takes place at distance $R_{\rm diss}\sim 0.3-1$ pc,
the bulk Lorentz factor of the jet is $\Gamma_j\sim
20$, and the size of the reconnection region $l'\simless 10^{16}$ cm. 
These quantities point towards an interesting possibility for the 
magnetic field structure in the jet and the origin of 
the blazar emission. If the magnetic field configuration is not
exactly axisymmetric at the horizon, the jet may emerge with
small-scale field structures of size similar to that of the 
central engine $\sim R_{\rm Sch}\sim 3\times
10^{14}M_9$ cm (along the direction of the jet propagation).  Even a modest
non-axisymmetry at the base of the jet can be amplified by stretching 
in the toroidal direction because of the jet (lateral)
 expansion. The jet expands from a lateral size of $r\sim R_{\rm Sch}$
 at the launching radius to $r=\theta_{\rm j}R_{\rm diss}\gg R_{\rm
   Sch}$ at the dissipation distance. The resulting scale of the
field reversals in the rest frame of the jet is  $\sim \Gamma_j R_{\rm
  Sch}$ that may be used as an estimate of the characteristic scale of the
reconnection layer $l'\simless 10^{16}$cm. For typical parameters, the 
reconnection time catches up with the expansion time of the jet
at distance $R_{\rm diss}\sim \Gamma_{\rm j}^2R_{\rm Sch}/\epsilon\sim 1$pc.

The very-high energy flares are modeled to be result of SSC or EIC
process of energetic electrons in the plasmoids, 
depending on the source. For the physical parameters inferred 
at the emitting region, however, the synchrotron emission from the
same population of electrons naturally peaks in the X-ray band.
In, at least some, of the sources fast $\gamma$-ray flares should also
be accompanied by fast X-ray flaring. Evidence for ultra-fast flares has
existed for some time. For instance Cui (2004) shows, using {\it RXTE}
observations of Mrk 421, that X-ray flaring on timescales as short as 
$\sim 10$ minutes is evident. Also, a characteristic envelope-fast
flares structure is evident (see their Fig.~10). Simultaneous 
detection of fast flares in both X-ray and $\gamma$-ray bands will be very
informative and constraining for the models.   

The bulk motion of plasmoids in the jet rest frame required for the model to work are 
very modest $\gamma_{\rm p}\sim 2$. The bulk motion in the reconnection picture
corresponds to the Alfv\'en speed: $\gamma_{\rm p}\simeq \sqrt{1+\sigma}$, implying
that a magnetization $\sigma\sim 3$ is required at the dissipation zone $R_{\rm diss}$.
Is it reasonable that the jet remains modestly Poynting-flux dominated at $\sim 1$pc
scale? This would imply that the conversion of Poynting-to-kinetic flux 
is not complete by that distance. A systematic study of superluminal motions
on pc to tens of pc scales reveals that blazar jets still undergo acceleration
out to the larger scale (Piner et al. 2012). In the context of MHD acceleration
 of jets, this would imply that, indeed, the pc-scale jet maintains a substantial
magnetization.

Ultra-fast flares are the tip of the iceberg of blazar variability.
The process of magnetic reconnection is potentially responsible for
powering a much broader range of blazar activity. Reconnection
may well take place at larger (e.g. multi pc) scale where the plasma
is presumably less magnetized (because of further conversion
of magnetic energy into kinetic). When the reconnecting plasma is characterized by 
$\sigma\simless 1$, the reconnection speed $v_{\rm rec}$ slows down (since
$v_{\rm rec}\propto V_A<c$). In this case, the reconnection timescale
becomes longer and reconnection layers may power days-to-weeks 
long flares of ``envelope'' emission.

\subsubsection{Other Astrophysical implications}

These ideas of plasmoid-dominated reconnection may be applied to other Astrophysical
sources. The variability patterns of gamma-ray burst (GRB) emission show fast flaring on top of 
slower-evolving lightcurves that may be connected to such reconnection process
(see also Lyutikov 2006; Lazar et
al. 2009; McKinney \& Uzdensky
2012; Zhang \& Yan 2011). Similar considerations
may apply to flares observed during the GRB afterglows (Giannios 2006).   
Reconnection minijets may also be the key to understand the fast GeV
flaring of the pulsar wind nebula of Crab (Clausen-Brown \& Lyutikov
2012; Cerutti et al. 2012).
In particular such model can attempt to explain the day-long
flares and shorter timescale variability observed during major
flaring of the Crab (Abdo et al. 2011; Tavani et al. 2011; Buehler et al. 2012).

\subsubsection{Open issues}

This study focused on the rough energetics and timescales of plasmoid-dominated
reconnection in blazar jets. While the feasibility of the process to account for blazar flares
has been made, a more detailed comparison to observations requires
progress in our theoretical understanding on a number of fronts. 

Where is the dissipation zone of jets located?
Studies of the global jet structure and stability can reveal 
where and why reconnection in the jet develops. These studies will
also probe the characteristic length scales and orientation of the
reconnection regions. The plasmoid-dominated reconnection 
is also a study in progress. Better understanding of fragmentation 
instabilities of the current sheet requires high-resolution 3D simulations. 
The theory should be tested and extended to the,
interesting here, trans-relativistic $\sigma \sim$a few regime.
Finally for making predictions on the spectra of the resulting
emission and direct comparison to observations, a better understanding
of particle acceleration in reconnection regions is
required. Particle-in-cell simulations make rapid progress in this
direction (Zenitani \& Hoshino 2005; Drake et al. 2006; Sironi \&
Spitkovsky 2011). 

\section*{Acknowledgments}

I thank H. Spruit and D. Uzdensky for insightful discussions and comments during the preparation of
the manuscript. I thank the referee for carefully reading the
manuscript, spotting inaccuracies in the derivations and making
suggestions that greatly improved the manuscript.

\appendix

\section{Properties of the reconnection region in blazars}

Here I estimate the Lundquist number $S=l'V_A/\eta$ that characterizes the reconnection
region in a blazar jet as well as several physical scales of relevance to the structure
of the current sheet. I focus on modestly high-$\sigma$ upstream (i.e., $\sigma\sim$ a few).
The resulting Alfv\'en speed $V_A=\sqrt{\sigma/(1+\sigma)}c\simeq
c$. The size of the reconnection region is constrained
from the observed duration of the blazar flares to $l'\simless 10^{16}$ cm. 

The resistivity $\eta$ contains contributions from Coulomb collisions $\eta_s$ (Spitzer resistivity)  
and electron scattering by photons  $\eta_C$  (the, so called,
``Compton drag''; Goodman \& Uzdensky 2008). For any reasonable 
temperature of the electrons in the jet, the Compton drag is expected to dominate over
the Spitzer resistivity. Following Goodman \& Uzdensky (2008)
$\eta_C=c\sigma_{\rm T}U'_{\rm rad}/3\pi e^2n_{\rm e}$.
For the very rough, order-of-magnitude estimate performed here we can set $U'_{\rm
  rad}\sim U'_{\rm j}$ (i.e., much of the energy density of the jet
-as given in eq.~(1)- 
converts to radiation) and $n_{\rm e}=n_{\rm p}=U'_{\rm j}/(1+\sigma)
m_pc^2$ for an electron-proton jet. For $R_{\rm diss}\sim 1$pc the resulting resistivity $\eta_C\sim 10^5$ 
cm$^2$/s. The Lundquist number is $S\sim 10^{21}$ clearly far in excess the critical value $S_c\simeq 10^4$
above which the secondary tearing instability in the current sheet
sets in (and $S\gg 10^5$ as required for monster plasmoids to develop). 
Reconnection is blazars is, therefore, likely to take place in the plasmoid-dominated
regime.
 
As discussed in Uzdensky et al. (2010) the plasmoids 
have a hierarchical structure with a large range in size. Large plasmoids are separated by a 
secondary reconnection layer of length $l^{(2)}$ where, 
in turn, smaller plasmoids form. This hierarchy repeats on ever
smaller scales $l^{(n)}$.
The smallest structure would formally be a ``critical layer'' of length $l_c=S_c\eta/c\sim 1$ mm with nominal
thickness $\delta_c=S_c^{-1/2}l_c\sim 10^{-3}$ cm! Such small structures are never realized in practice 
because the characteristic plasma scales (e.g. the ion skin depth $d_i$) are much larger. 
For the expected plasma density of the jet $d_i\sim 10^8$ cm. 
Once the thickness of a sublayer $\delta \sim d_i$, the resistive MHD description fails.
One instead deals with ``collisionless'' reconnection where the
resistivity is likely to be controlled by plasma
instabilities.  In this regime, the Petschek model for (relativistic) reconnection (Lyubarsky 2005)
may apply (Kulsrud 2001). The Petschek reconnection rate at a layer of
length $l=S_c^{1/2}d_i\sim 10^{10}$ cm is $v_{\rm rec}\sim c/\ln S_p \sim 0.03 c$, where $S_p=lc/\eta\sim 10^{15}$.          
Moreover, when $\delta \sim d_i$, Hall MHD terms become important
further increasing the reconnection rate (see, e.g., Malyshkin 2008).
This local fast reconnection rate controls also the overall rate of
large-scale reconnection (Uzdensky et al. 2010).

\end{document}